\documentclass[conference,twoside]{IEEEtran}
\usepackage[dvips]{epsfig}
\usepackage{cite,algorithm,algorithmic,amsmath,amssymb, amsthm}
\usepackage[dvips]{graphicx}
\usepackage{subfigure,amsfonts,balance}
\usepackage{epstopdf}

\DeclareMathOperator{\tr}{\mathrm{tr}}
\DeclareMathOperator{\rank}{\mathrm{rank}}
\DeclareMathOperator{\E}{\mathbb{E}}

\theoremstyle{remark}
\newtheorem{remark}{Remark}

\begin{document}
%
\title{Wiretapped Signal Leakage Minimization \\for Secure Multiuser MIMO Systems \\via Interference Alignment}


\author{
\IEEEauthorblockN{Tung T. Vu$^{1,2}$, Ha Hoang Kha$^2$, Trung Q. Duong$^3$, Nguyen-Son Vo$^1$}

\IEEEauthorblockA{\small
$^1$ Duy Tan University, Vietnam (e-mail: \{vuthanhtung1,vonguyenson\}@dtu.edu.vn) \\
$^2$ Ho Chi Minh City University of Technology, Vietnam (e-mail: hhkha@hcmut.edu.vn) \\
$^3$ Queen's University Belfast, UK (e-mail: trung.q.duong@qub.ac.uk) \\
    }
        \thanks{}
    \thanks{}
}
\maketitle
\begin{abstract}
\boldmath
In this paper, interference alignment (IA) is designed for secure multiuser multiple-input multiple-output systems in the presence of an eavesdropper. The proposed IA technique designs the transmit precoding and receiving subspace matrices to minimize both the total inter-main-link interference and the wiretapped signals. With perfect channel state information is known at the transmitters (Txs), the cost function of the optimization problem is alternatively minimized over the precoding matrices at the Txs and the receiving subspace matrices at the receivers (Rxs) and the eavesdropper. The feasible condition, the proof of convergence of the proposed IA approach are provided. Numerical results reveal a significant improvement in terms of average secrecy sum rate of our IA algorithm as compared to the conventional IA design.
\end{abstract}
%

\IEEEpeerreviewmaketitle
\section{Introduction}
\label{sec:introd}
Due to the broadcasting nature of the wireless multiuser multiple-input multiple-output (MIMO) communication networks, the data transmission among legitimate users is susceptible to be wiretapped by nearby eavesdroppers. Various signal processing techniques in the physical layer have been developed to enhance secrecy against the eavesdropper \cite{Hanif14,Yang14,Geraci12,Yang13,Mukherjee09}. However, all the solutions in \cite{Yang13,Hanif14,Yang14,Geraci12,Mukherjee09} mainly focuses on the optimization problems of secrecy sum rate (SSR) in the broadcast channels in which only one base station communicates with multiple users in the network. The secure communication for multiple transmitter-receiver (Tx-Rx) pair transmissions has not been studied. It should be emphasized that the sum rate maximization of the MIMO interference channels with multiple user pairs is mathematically challenging even for the scenarios without taking secure transmission into consideration. Interference alignment (IA) has recently been exploited as a potential technique to increase the secrecy in  multiuser MIMO systems \cite{Koyl11,Sasaki12}. The authors in \cite{Koyl11} has proven that it is possible for each user in the network to achieve a nonzero secure degrees of freedom (DoF) when using an IA scheme along with precoding matrix at each Tx. It has been shown in \cite{Sasaki12} that the secure transmission is also feasible where the number of antennas at legitimate Tx ans Rx is greater than those of the eavesdropper. However, a detailed IA design for secure multisuer MIMO communication networks has not been explored in the literature.

In this paper, we study on the secure communication for multiuser MIMO systems in presence of an eavesdropper. Inspired by the standard IA technique in \cite{cadambe08,kumar10,peters09}, we develop the IA scheme, namely the wiretapped-signal-leakage-minimization (WSLM) IA technique, for the secrecy transmission in multiuser MIMO systems.
By forcing the receiving signals at the eavesdropper into reduced-dimensional receiving subspace, and then minimizing the total signal power, the desired signal will become harder to be recoverd at the eavesdropper. The standard IA technique has been well studied in MIMO systems without the eavesdropper \cite{cadambe08,kumar10,peters09}. The authors in \cite{cadambe08} have proposed a scheme which make the Rxs interference-free by forcing interfering signals into a reduced-dimensional subspace. However, closed-form solutions of the precoders with more than 3 users appear to be critical. Therefore, iterative techniques have been investigated \cite{kumar10,peters09}, in which the precoding matrices at Txs and the interference subspace matrices at Rxs are updated in every iteration.
Compared with the scheme in \cite{cadambe08}, the iterative algorithm can efficiently obtain the optimal precoding matrices and interference subspaces for the system with more than 3 Tx-Rx pairs. Motivated by these works, we adopt the iterative approach to design IA for the secure multiuser MIMO systems. We also derived the proof of convergence, the feasible condition and the analysis on system parameters for the best IA design. Simulation results show that the proposed IA design outperforms the conventional approach \cite{peters09} in terms of average SSR.

The rest of this paper is organized as follows. Section \ref{sec:Model} introduces the multiuser MIMO system in the presence of an eavesdropper. In Section \ref{Sec:IAMUMIMO}, we propose the IA algorithm. The numerical results are provided in Section \ref{sec:Results}. Finally, Section \ref{sec:Conclusion} presents concluding remarks.

\emph{Notation}: $\pmb{X}$ and $\pmb{x}$ are respectively denoted as a complex matrix and a vector. $\pmb{X}^T$ and $\pmb{X}^H$ are the transposition and conjugate transposition of the complex matrix $\pmb{X}$, respectively. $\pmb{I}$ and $\pmb{0}$ are respectively identity and zero matrices with the appropriate dimensions. $\tr (.)$, $\rank (.)$ and $\E(.)$  are the trace, rank and expectation operators, respectively. $||\pmb{X}||_F$ is the Frobenius norm. $\pmb{x}\sim \mathcal{CN}(\bar{\pmb{x}},\pmb{R}_{\pmb{x}})$ means that $\pmb{x}$ is a complex Gaussian random vector with means $\bar{\pmb{x}}$ and covariance $\pmb{R}_{\pmb{x}}$. $\mathcal{R}(\pmb{X})$ and $\mathcal{N}(\pmb{X})$ are respectively the range and nullity of matrix $\pmb{X}$. $\mathcal{X}^\bot$ and $\mathrm{dim}\{\mathcal{X}\}$ is denoted as the orthogonal subspace and the dimension of subspace $\mathcal{X}$, respectively.


\section{System Model}
\label{sec:Model}
\begin{figure}[h!]
\begin{center}
\epsfxsize=7.5cm
\leavevmode\epsfbox{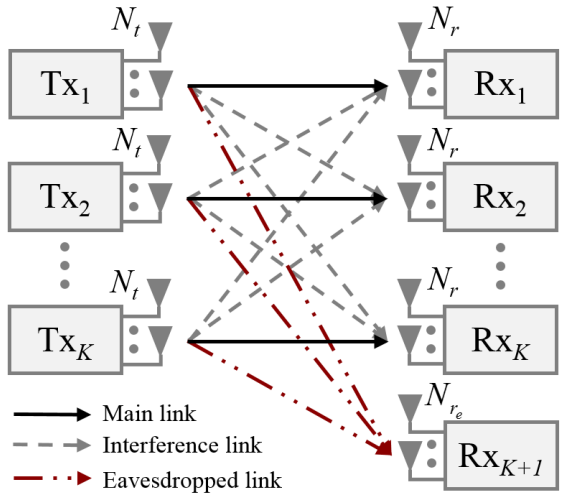}
\caption{A system model of secure multiuser MIMO communication network.}
\label{fig:DownlinkMIMOIC}
\end{center}
\end{figure}
Consider a MIMO interference channel with $K$ Tx-Rx pairs and an eavesdropper as depicted in Fig. \ref{fig:DownlinkMIMOIC}. Without loss of generality, we assume that the eavesdropper is the $(K+1)$-th Rx with the $N_{r_e}$ antennas, all Txs and Rxs are equipped with the same number of antennas $N_t$ and $N_r$, respectively, and communicate with the same number of data stream $d$.  The received signal $\pmb{y}_k \in \mathbb{C}^{N_r \times 1}$ at the $k$-th Rx for $k \in \mathcal{K}=\{1,...,K\}$ can be given as
\begin{align}\label{Signal:Rx}
\pmb{y}_k = \sum\limits_{\ell = 1}^K \pmb{H}_{k,\ell}\pmb{F}_{\ell}\pmb{s}_\ell  +\pmb{n}_k
\end{align}
where $\pmb{H}_{k,\ell} \in \mathbb{C}^{N_r \times N_t}$ is the static flat-fading MIMO channels matrix from the $\ell$-th Tx to the $k$-th Rx, $\pmb{F}_k \in \mathbb{C}^{N_t \times d}$ is the precoding matrix applied on the signal vector $\pmb{s}_k \in \mathbb{C}^{d \times 1}$ and $\pmb{n}_k \sim \mathcal{CN}\left( 0,\sigma _k^2\pmb{I}_{N_r} \right)$ is a complex Gaussian noise vector. Assuming that $\E\{ {{\pmb{s}_k}\pmb{s}_k^H} \} = {\pmb{I}_{{d}}}$ and the global channel state information (CSI) is available, the channel capacity at the $k$-th legitimate Rx, for $k \in \mathcal{K}$, can be calculated as \cite{Bazz12}
\begin{align}\label{Cap:Rxk}
\mathcal{R}_k = \log _2\left| \pmb{I}_{N_r} + \pmb{H}_{k,k}\pmb{F}_k\pmb{F}_k^H\pmb{H}_{k,k}^H\pmb{R}_{z_k}^{-1} \right|
\end{align}
where $\pmb{R}_{z_k} = \sum\limits_{\ell = 1,\ell \ne k}^K \pmb{H}_{k,\ell}{\pmb{F}_\ell\pmb{F}_\ell^H\pmb{H}_{k,\ell}^H}  + \sigma _k^2\pmb{I}_{N_r}$ is the correlation matrix of interference and noise in Eq. \eqref{Signal:Rx}. The eavesdropper or the $(K+1)$-th Rx wiretaps the data signal from the $k$-th Tx-Rx pair and the information leakage rate of this wiretap channel can be given by
\begin{align}\label{Cap:WiretapRxk}
\mathcal{R}_k^{(e)} = \log _2\left| \pmb{I}_{N_{r_e}} + \pmb{H}_{K+1,k}\pmb{F}_k\pmb{F}_k^H\pmb{H}_{K+1,k}^H\pmb{R}_{e,k}^{-1} \right|
\end{align}
where $\pmb{R}_{e,k} = \sum\limits_{\ell = 1,\ell \ne k}^K \pmb{H}_{K+1,\ell}{\pmb{F}_\ell\pmb{F}_\ell^H\pmb{H}_{K+1,\ell}^H}  + \sigma _k^2\pmb{I}_{N_{r_e}}$ is the interference plus noise correlation matrix at the eavesdropper. Now, the secrecy rate for the $k$-th Tx-Rx pair can be defined as
\begin{align}\label{Cap:SRRxk}
\mathcal{R}_{S,k}  = [\mathcal{R}_k-\mathcal{R}_k^{(e)}]^{+}.
\end{align}
Finally, the SSR of multiuser MIMO system can be expressed as
\begin{align}\label{SSR}
\mathcal{R}_{S} = \sum\limits_{k = 1}^K\mathcal{R}_{S,k} = \sum\limits_{k = 1}^K [\mathcal{R}_k-\mathcal{R}_k^{(e)}]^{+},
\end{align}
where $[a]^{+}=\max(a,0)$.

\section{WSLM for Secrecy Multiuser MIMO Communication via IA}
\label{Sec:IAMUMIMO}
The aim of WSLM IA design is to find the precoding matrices $\{\pmb{F}_k\}_{k=1}^K$  which aligns all interference signals at the legitimate Rxs into reduced subspaces and all signals at the eavesdropper  into a subspace. Mathematically, the precoder matrices $\{\pmb{F}_k\}_{k=1}^K$ and the receiving signal subspace matrix $\{\pmb{W}_k\}_{k=1}^{K+1}$ (at the legitimate Rxs and the eavesdropper) satisfy the three following conditions
\begin{eqnarray}\label{IA:SecrecyConditions1}
  \mathrm{rank}\big(\pmb{W}_k^H\pmb{H}_{k,k}\pmb{F}_k\big) &=& d  \\\label{IA:SecrecyConditions2}
  \pmb{W}_k^H\pmb{H}_{k,\ell}\pmb{F}_\ell &=& \mathbf{0};\,\, \forall \ell \neq k, \ell \in \mathcal{K}\\\label{IA:SecrecyConditions3}
  \pmb{W}_{K+1}^H\pmb{H}_{K+1,\ell}\pmb{F}_\ell &=& \mathbf{0}; \,\, \forall \ell \in \mathcal{K}.
\end{eqnarray}
Conditions \eqref{IA:SecrecyConditions1} guarantees that the intended signal at the $k$-th Rx achieves $d$ degree of freedom when $\pmb{H}_{k,k}$ is full rank. The condition \eqref{IA:SecrecyConditions2} ensures no inter-main link interference (IMLI), i.e., no interference among the legitimate users, at the $k$-th Rx. Condition \eqref{IA:SecrecyConditions3} makes certain that there is no leakage signal wiretapped at the eavesdropper or the $(K+1)$-th Rx.

To remove the IMLIs at the intended $k$-th Rx, we align the interference signal $\pmb{H}_{k,\ell}\pmb{F}_\ell$ into the interference receiving subspace $\mathcal{W}_k^{\perp}$ which is spanned by the orthonormal basis matrix $\pmb{U}_k$. Hence, the total interference leakage inside the receiving subspace is defined by
\begin{eqnarray} \label{J1:2}
 &&\mathfrak{J}_1(\{\pmb{F}_k\}_{k=1}^K, \{\pmb{U}_k\}_{k=1}^K) \\ \nonumber
 &&= \sum\limits_{k=1}^{K}\sum\limits_{\ell=1,\ell \neq k}^{K} \label{J1}
 ||\pmb{H}_{k,\ell}\pmb{F}_\ell - \pmb{U}_k\pmb{U}_k^H\pmb{H}_{k,\ell}\pmb{F}_\ell ||_\emph{F}^2.
\end{eqnarray}

To reduce the wiretapped information, we align all the receiving signals at the eavesdropper into a receiving subspace $\mathcal{W}_{K+1}$ spanned by the orthonormal basis matrix $\pmb{W}_{K+1}$ where their total signal power is minimized. Mathematically, we aim to minimize
\begin{align} \label{J2:2}
 &\mathfrak{J}_2(\{\pmb{F}_k\}_{k=1}^K, \pmb{U}_{K+1}) \\ \nonumber
 &= \sum\limits_{k=1}^{K}||\pmb{H}_{K+1,\ell}\pmb{F}_\ell - \pmb{U}_{K+1}\pmb{U}_{K+1}^H\pmb{H}_{K+1,\ell}\pmb{F}_\ell ||_\emph{F}^2
\end{align}
where $\pmb{U}_{K+1}$ is the orthonormal basis of the subspace $\mathcal{W}_{K+1}^\bot$.

Finally, the IA problem can be expressed as the following joint optimization problem
\begin{subequations}\label{OptProIA1}
\begin{eqnarray}
\underset{\{\pmb{F}_k\}_{k=1}^K,\{\pmb{U}_k\}_{k=1}^{K+1}}{\text{min}} ~~&&\mathfrak{J}(\{\pmb{F}_k\}_{k=1}^K, \{\pmb{U}_k\}_{k=1}^{K+1}) \\
\text{s.t.} \qquad &&\pmb{F}_k^H\pmb{F}_k = \frac{P_t}{d}\pmb{I}; \,\,\, \forall k \in \mathcal{K} \qquad \\
 \qquad  &&\pmb{U}_k^H\pmb{U}_k = \pmb{I};  \forall k \in \mathcal{K}\cup\{K+1\} \qquad
\end{eqnarray}
\end{subequations}
where
\begin{align}\label{J}
&\mathfrak{J}(\{\pmb{F}_k\}_{k=1}^K, \{\pmb{U}_k\}_{k=1}^{K+1}) \\ \nonumber
&=  \mathfrak{J}_1(\{\pmb{F}_k\}_{k=1}^K, \{\pmb{U}_k\}_{k=1}^K)+\mathfrak{J}_2(\{\pmb{F}_k\}_{k=1}^K, \pmb{U}_{K+1})
\end{align}
We now derive the solution of problem \eqref{OptProIA1} via alternating minimization \cite{peters09} in the following.

\subsubsection*{Transmit precoder selection}
When $\{\pmb{U}_k\}_{k=1}^{K+1}$ are fixed, for finding $\pmb{F}_\ell$, \eqref{OptProIA1} reduces to the optimization problem
\begin{subequations}\label{OptProF:IA1}
\begin{align}
 &\underset{\pmb{F}_\ell}{\text{min}} \qquad \mathrm{tr}\left\{\pmb{F}_\ell^H\left(\sum\limits_{k=1,k \neq \ell}^{K+1}\pmb{H}_{k,\ell}^H(\pmb{I}- \pmb{U}_k\pmb{U}_k^H)\pmb{H}_{k,\ell}\right)\pmb{F}_\ell\right\}\\
 &\text{s.t.} \qquad \pmb{F}_\ell^H\pmb{F}_\ell = \frac{P_t}{d}\pmb{I}, \,\,\, \forall \ell \in \mathcal{K}
\end{align}
\end{subequations}
The solution to the above optimization problem is given by \cite{Lutke97}
\begin{align}\label{Fl:IA1}
  \pmb{F}_\ell = \sqrt{\frac{P_t}{d}}\zeta_{\min}^d\left\{\sum\limits_{k=1,k \neq \ell}^{K+1}\pmb{H}_{k,\ell}^H(\pmb{I}- \pmb{U}_k\pmb{U}_k^H)\pmb{H}_{k,\ell}\right\}
\end{align}
where $\zeta_{\min}^d\left\{\pmb{X}\right\}$ is a matrix whose columns are the $d$ eigenvectors corresponding to the $d$ smallest eigenvalues of $\pmb{X}$.

\subsubsection*{Receiver interference subspace selection}
Secondly, when $\pmb{F}_\ell$ is fixed,
from \eqref{OptProIA1}, the orthonormal basis $\pmb{U}_k$ can be selected by solving the following optimization problem
\begin{subequations}\label{ConventionalOptProUk}
\begin{eqnarray}
 \underset{\pmb{U}_k}{\text{max}} && \mathrm{tr}\left\{\pmb{U}_k^H\left(\sum\limits_{\ell=1,\ell \neq k}^{K}\pmb{H}_{k,\ell}\pmb{F}_\ell\pmb{F}_\ell^H\pmb{H}_{k,\ell}^H\right)\pmb{U}_k\right\} \\
 \text{s.t.} && \pmb{U}_k^H\pmb{U}_k = \pmb{I}.
\end{eqnarray}
\end{subequations}
Hence, $\pmb{U}_k$ can be given by \cite{Lutke97}
\begin{align}\label{ConventionalUk}
  \pmb{U}_k = \zeta_{\max}^{N_r-d}\left\{\sum\limits_{\ell=1,\ell \neq k}^{K}\pmb{H}_{k,\ell}\pmb{F}_\ell\pmb{F}_\ell^H\pmb{H}_{k,\ell}^H\right\}
\end{align}
where $\zeta_{\max}^{N_r-d}\left\{\pmb{X}\right\}$ is a matrix whose columns are the $(N_r-d)$ dominant eigenvectors of matrix $\pmb{X}$.

To deal with wiretapped signals, we assume that $\pmb{W}_{K+1}$ is a $N_{r_e}\times n$ matrix where $n\leq N_{r_e}$. Since $\pmb{W}_{K+1}$ is an orthonomal basis matrix, $\mathrm{rank}\left(\pmb{W}_{K+1}\right)=\mathrm{rank}\left(\pmb{W}_{K+1}^H\right)=n$. To satisfy condition \eqref{IA:SecrecyConditions3}, $\pmb{H}_{K+1,\ell}\pmb{F}_\ell \in \mathbb{C}^{N_{r_e}\times d}$ must lie in the null space of $\pmb{W}_{K+1}^H$. Hence, the existence condition of $\pmb{W}_{K+1}^H$ is $N_{r_e}-n\geq d$. Thus, we consider the case that $n=N_{r_e}-d$ for the dimension of the null space of $\pmb{W}_{K+1}^H$ to be smallest, to make the eavesdropper harder to recover signal. Then, the matrix size of $\pmb{U}_{K+1}$ must be $N_{r_e}\times d$. From \eqref{OptProIA1}, the orthonormal basis $\pmb{U}_{K+1}$ can be found by solving the following optimization problem
\begin{subequations}\label{OptProUe}
\begin{align}
 \underset{\pmb{U}_{K+1}}{\text{max}} &\qquad \mathrm{tr}\left\{\pmb{U}_{K+1}^H\left(\sum\limits_{\ell=1}^{K}\pmb{H}_{K+1,\ell}\pmb{F}_\ell\pmb{F}_\ell^H\pmb{H}_{K+1,\ell}^H\right)\pmb{U}_{K+1}\right\} \\
 \text{s.t.} &\qquad \pmb{U}_{K+1}^H\pmb{U}_{K+1} = \pmb{I}_d.
\end{align}
\end{subequations}
The solution to \eqref{OptProUe} can be found as follows \cite{Lutke97}
\begin{align}\label{Ue}
  \pmb{U}_{K+1} = \zeta_{\max}^d\left\{\sum\limits_{\ell=1}^{K}\pmb{H}_{K+1,\ell}\pmb{F}_\ell\pmb{F}_\ell^H\pmb{H}_{K+1,\ell}^H\right\}
\end{align}
The step-by-step iterative algorithm for the WSLM method can be shown in Algorithm \ref{alg1}.

\begin{algorithm}[ht]
\caption{: Proposed  WSLM IA Algorithm for Secure Multiuser MIMO Systems}\label{IterIA}
\begin{algorithmic}[1]\label{alg1}
\STATE Inputs: $d,\pmb{H}_{k,\ell}$,  $\forall k \in \mathcal{K}\cup\{K+1\}$, $\forall \ell \in \mathcal{K}$, $\kappa=0$, $\kappa_{\max }$, where $\kappa$ is the iteration index;
\STATE Initial variables: random matrix $\{\pmb{F}_k^{(0)}\}_{k = 1}^{K}$ satisfied $\pmb{F}_k^{(0)H}\pmb{F}_k^{(0)}=\sqrt{\frac{P_t}{d}}\pmb{I}_d$; then select $\{\pmb{U}_k^{(0)}\}_{k = 1}^{K+1}$ from \eqref{ConventionalUk} and \eqref{Ue};
\STATE Evaluate the objective function $\mathfrak{J}(\{\pmb{F}_k^{(0)}\}_{k=1}^K, \{\pmb{U}_k^{(0)}\}_{k=1}^{K+1})$ from \eqref{OptProIA1};
\WHILE{$\kappa < \kappa_{\max}$}
\STATE For fixed $\{\pmb{U}_k^{(\kappa)}\}_{k = 1}^{K+1}$, select $\{\pmb{F}_k^{(\kappa+1)}\}_{k = 1}^K$ from \eqref{Fl:IA1};
\STATE For fixed $\{\pmb{F}_k^{(\kappa+1)}\}_{k = 1}^K$, select $\{\pmb{U}_k^{(\kappa+1)}\}_{k = 1}^{K+1}$ from \eqref{ConventionalUk} and \eqref{Ue};
\STATE Evaluate the objective function $\mathfrak{J}(\{\pmb{F}_k^{(\kappa+1)}\}_{k=1}^K, \{\pmb{U}_k^{(\kappa+1)}\}_{k=1}^{K+1})$ from \eqref{OptProIA1};
\STATE $\kappa=\kappa+1$;
\STATE Repeat steps 5-8 until convergence.
\ENDWHILE
\end{algorithmic}
\end{algorithm}

\subsection{Convergence analysis for the WSLM IA design}
In the $\kappa$-th iteration, after step 5 and 6 in Algorithm \ref{alg1}, it can be verified that
 \begin{equation}
\mathfrak{J}(\{\pmb{F}_k^{(\kappa+1)}\}_{k=1}^K, \{\pmb{U}_k^{(\kappa+1)}\}_{k=1}^{K+1}) \leq \mathfrak{J}(\{\pmb{F}_k^{(\kappa)}\}_{k=1}^K, \{\pmb{U}_k^{(\kappa)}\}_{k=1}^{K+1})
\end{equation}
That is, the cost function $\mathfrak{J}$ in \eqref{OptProIA1} is reduced monotonically over iteration. In addtion, the cost function is bounded below by zero. Thus, the convergence of the first proposed IA algorithm is guaranteed.

\subsection{Feasible condition for the proposed WSLM IA design}
The feasibility of the IA scheme is well studied in \cite{yetis09} but not in the security context and, thus, in this subsection, we derive the feasible condition for the proposed IA approach. It has been shown in \cite{yetis09} that the proper systems surely render the feasibility of the IA problems. The MIMO interference channels are known as proper systems if the number of variables $N_v$ is greater than or equal to the number of equations $N_{eq}$ in the IA conditions.
The total number of equations $N_{eq}$ is directly given from \eqref{IA:SecrecyConditions2} and \eqref{IA:SecrecyConditions3} as follows
\begin{align}\label{Neq}
  N_{eq} &= K(K-1)d^2+K(N_{r_e}-d)d.
\end{align}
The number of variables to be designed for the precoder $\pmb{F}_\ell$, subspace matrices $\pmb{W}_k$ and $\pmb{W}_{K+1}$ equals $d(N_t-d)$, $d(N_r-d)$ and $d(N_{r_e}-d)$, respectively \cite{yetis09}. Finally, the total number of variables in the proposed IA conditions \eqref{IA:SecrecyConditions2} and \eqref{IA:SecrecyConditions3} can be given as
\begin{align}\label{Nv}
  N_v = Kd\left(N_t+N_r-2d\right)+d(N_{r_e}-d).
\end{align}

Denote $\left(N_t\times N_r,N_{r_e},d\right)^K$ as the system where there are $K$ Tx-Rx pairs and one eavesdropper; the Tx, Rx and eavesdropper are respectively equipped with $N_t$, $N_r$, $N_{r_e}$ antennas; and $d$ data streams are sent between each Tx-Rx pair.
By comparing $N_{eq}$ in \eqref{Neq}  and $N_v$ in \eqref{Nv}, the IA scheme in Algorithm \ref{alg1} for the $\left(N_t\times N_r,N_{r_e},d\right)^K$ system with secure communication is feasible if $N_v\geq N_{eq}$, i.e., \begin{align}\label{feasibleIA1condition}
  K(N_t+N_r)-\left(K^2+1\right)d \geq N_{r_e}(K-1).
\end{align}
%
\begin{remark}
Condition \eqref{feasibleIA1condition} implies that for satisfying the WSLM IA feasible conditions, the number of antennas at the eavesdropper is restricted by
$N_{r_e}\leq\frac{K(N_t+N_r)-\left(K^2+1\right)d}{K-1}$. Since the number of antennas at the eavesdropper cannot be controlled in the proposed system, the secrecy can be enhanced when increasing the number of transceiver antennas $N_t$ and $N_r$.
\end{remark}
\subsection{Relationship between information-leakage-rate and the number of antennas at the eavesdropper}
In the WSLM IA design, we aim to minimize the total wiretapped signal power in a reduced dimension subspace and, thus, the sum information-leakage-rate (SILR), defined as $\displaystyle \sum_{k=1}^K \mathcal{R}_k^{(e)}$, can be decreased. However, increasing $N_{r_e}$ means increasing the dimension of the decoding space at the eavesdropper and, therefore, results in an increase in the SILR. Hence, one can ask which is the relationship between $N_{r_e}$ and the SILR in the WSLM IA design. We investigate the answer of this question in this subsection.
From \eqref{IA:SecrecyConditions2}, we have
\begin{eqnarray}\label{rangeFell}
\mathcal{R}(\pmb{F}_\ell) &\subseteq& \mathcal{R}(\pmb{U}_k^H\pmb{H}_{k,\ell}), \forall \ell \neq k, \ell \in \mathcal{K} \\
\mathcal{R}(\pmb{F}_k) &\subseteq& \mathcal{N}(\pmb{U}_k^H\pmb{H}_{k,\ell}), \forall \ell \neq k, \ell \in \mathcal{K}.\label{rangeFk}
\end{eqnarray}
From \eqref{IA:SecrecyConditions3}, the eavesdropper can overhear the signal of the $k$-th Tx if the subspace spanned by $\pmb{F}_k$ lies in the subspace spanned by $\pmb{W}_{K+1}^H\pmb{H}_{K+1,k}$, i.e,
\begin{eqnarray}\label{rangeFkatE}
\mathcal{R}\left(\pmb{F}_k\right) &\subseteq&  \mathcal{R}\left(\pmb{W}_{K+1}^H\pmb{H}_{K+1,k}\right), \forall k \in \mathcal{K}.
\end{eqnarray}
It is noted that
\begin{eqnarray}\label{dimRangeUk}\nonumber
\mathrm{dim}\{\mathcal{R}\left(\pmb{U}_k^H\pmb{H}_{k,\ell}\right)\}&=&\mathrm{rank}\left(\pmb{U}_k^H\pmb{H}_{k,\ell}\right)\\
&=&\mathrm{rank}\left(\pmb{U}_k\right)=N_r-d,\\\nonumber\label{dimnullUk}
\mathrm{dim}\{\mathcal{N}\left(\pmb{U}_k^H\pmb{H}_{k,\ell}\right)\}&=& M - \mathrm{dim}\{\mathcal{R}\left(\pmb{U}_k^H\pmb{H}_{k,\ell}\right)\}\\
&=& N_t - N_r + d,\\ \nonumber
\mathrm{dim}\left\{\mathcal{R}\left(\pmb{W}_{K+1}^H\pmb{H}_{K+1,k}\right)\right\} &=& \mathrm{rank}\left(\pmb{W}_{K+1}^H\pmb{H}_{K+1,k}\right) \\
\nonumber
&=& \mathrm{rank}\left(\pmb{W}_{K+1}\right) \\
&=& N_{r_e}-d.
\end{eqnarray}
From \eqref{rangeFk} and \eqref{rangeFkatE}, it is implied that the eavesdropper can recover the signal of the $k$-th Tx easier if the dimension of space for recovering this desired signal at the eavesdropper is larger than the dimension of the space where this desired signal belongs to. Mathematically, the SILR will start increasing when
\begin{eqnarray}
  \mathrm{dim}\{\mathcal{R}\left(\pmb{W}_{K+1}^H\pmb{H}_{K+1,k}\right)\}&\geq& \mathrm{dim}\{\mathcal{N}\left(\pmb{U}_k^H\pmb{H}_{k,\ell}\right)\},\\ \label{dimWk}
  \Leftrightarrow N_{r_e} &\geq& N_t - N_r + 2d. \label{Ne:condition}
\end{eqnarray}
\begin{remark}
It can be seen from \eqref{dimRangeUk} and \eqref{dimnullUk} that if $N_r$ increases, $\mathrm{dim}\{\mathcal{R}\left(\pmb{U}_k^H\pmb{H}_{k,\ell}\right)\}$ increases and $\mathrm{dim}\{\mathcal{N}\left(\pmb{U}_k^H\pmb{H}_{k,\ell}\right)\}$ decreases, which makes $\pmb{F}_\ell$ and $\pmb{U}_k$ have the larger dimension of the search space and easier to reach the optimal solution, for all $\ell,k \in \mathcal{K}$. Hence, the main-link rate is increased when $N_r$ increases. In addition, it can be seen from \eqref{Ne:condition} that when $N_r$ increases, $(N_t-N_r+2d)$ decreases, which means the number of antennas required for keeping a low SILR at the eavesdropper becomes smaller. Therefore, there should be a trade-off between maximizing main-link-rate and the SILR for different values of $N_r$.
\end{remark}
\section{Simulation Results}
\label{sec:Results}
In this section, we evaluate the SSR performance of our proposed IA design through some numerical results in comparison with the conventional IA algorithm \cite{peters09}. In simulations, noise variances are normalized $\sigma_k^2=\sigma^2=1$. The Rayleigh fading channel coefficients are generated from the complex Gaussian distribution ${\mathcal{CN}}(0,1)$. We define signal-to-noise-ratio $\text{SNR}=\frac{P_{t}}{\sigma^2}$. All the numerical results are averaged over the $200$ channel realizations.

\begin{figure}[b!]
\begin{center}
\epsfxsize=7.5cm
\leavevmode\epsfbox{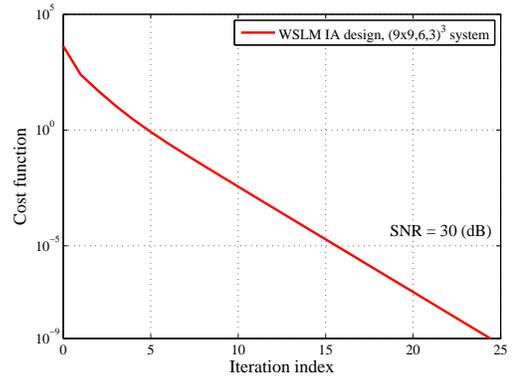}
\caption{The convergence behavior of the proposed IA algorithm.}
\label{fig:Convergence}
\end{center}
\end{figure}

First, we investigate the convergence characteristic of the WSLM IA algorithm.  We run the simulation for a ${\left( 9 \times 9,6,3 \right)^3}$ system with a random channel realization. Note that this system satisfies the feasible conditions \eqref{feasibleIA1condition}. The evolution of the cost functions (SSR) over iterations is illustrated in Fig. \ref{fig:Convergence}. As can been seen from Fig. \ref{fig:Convergence}, the cost function reduces monotonically over iteration and approaches to zero. The proposed IA algorithm takes around $25$ iterations to converge to $10^{-9}$.

\begin{figure}[t!]
\begin{center}
\epsfxsize=7.5cm
\leavevmode\epsfbox{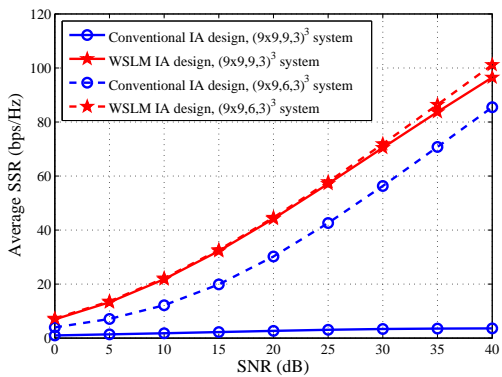}
\caption{The average SSR versus SNR for ${\left( 9 \times 9,6,3 \right)^3}$ and ${\left( 9 \times 9,9,3 \right)^3}$ systems.}
\label{fig2}
\end{center}
\end{figure}

Now, we compare the average SSR of our proposed IA algorithm with that of the conventional IA method \cite{peters09} which does not involve the secrecy constrains. We consider two scenarios of ${\left( 9 \times 9,6,3 \right)^3}$ and ${\left( 9 \times 9,9,3 \right)^3}$ systems.
It should be noted that the WSLM IA feasible condition is satisfied in these two systems. It has been revealed from Fig. \ref{fig2} that the WSLM IA scheme significantly improve the SSR in comparison with the conventional IA design when $N_{r_e}$ is high. That is because when $N_{r_e}$ is small, the SILR is small and hence, the difference between our IA design and the conventional IA one is not noticeable.
\begin{figure}[t!]
\begin{center}
\epsfxsize=7.5cm
\leavevmode\epsfbox{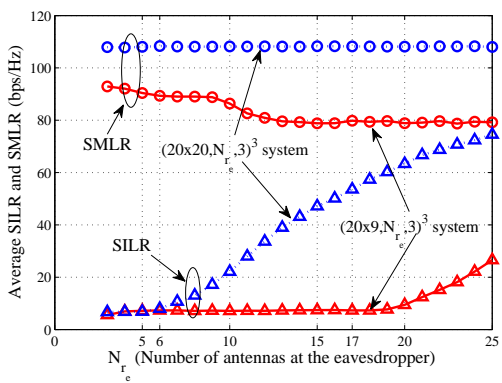}
\caption{The average SILR and SMLR versus $N_{r_e}$ for ${\left( 20 \times 9,N_{r_e},3 \right)^3}$ and ${\left( 20 \times 20,N_{r_e},3 \right)^3}$ systems.}
\label{fig3}
\end{center}
\end{figure}
\begin{figure}[h!]
\begin{center}
\epsfxsize=7.5cm
\leavevmode\epsfbox{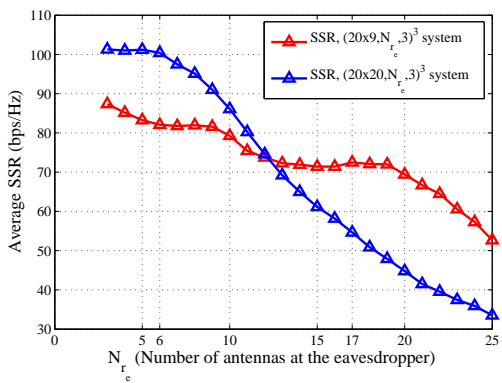}
\caption{The average SSR versus $N_{r_e}$ for ${\left( 20 \times 9,N_{r_e},3 \right)^3}$ and ${\left( 20 \times 20,N_{r_e},3 \right)^3}$ systems.}
\label{fig4}
\end{center}
\end{figure}

Finally, we investigate impacts of the number of receive antennas $N_{r_e}$ on the sum main-link-rate (SMLR) and the SILR. Fig. \ref{fig3} illustrates the average SMLR, SILM while Fig. \ref{fig4} plots the SSR of the proposed IA design for the ${\left( 20 \times 9,N_{r_e},3 \right)^3}$ and ${\left( 20 \times 20,N_{r_e},3 \right)^3}$ systems with $N_{r_e}=3,...,25$ at $SNR=30$ dB. As such parameters, the WSLM IA feasible condition is satisfied in these two systems. It should be noted from \eqref{Ne:condition} that the SILR will start increasing if $N_{r_e}\geq 6$ in the ${\left( 20 \times 20,N_{r_e},3 \right)^3}$ system and $N_{r_e}\geq 17$ in the ${\left( 20 \times 9,N_{r_e},3 \right)^3}$ system. These analytical results have been confirmed by the numerical simulation in Fig. \ref{fig3}. On the other hand, Figs. \ref{fig3} and \ref{fig4} show the trade-off between the SMLR and SILR for different number of receive antennas $\{N_r, N_{r_e}\}$.  When $N_r$ is large (e.g., $N_r=20$) the SMLR is higher. However, the number of antennas $N_{r_e}$ required for keeping a low SILR at the eavesdropper is smaller (see \eqref{Ne:condition}). Thus, the SSR for the case $N_r=20$ is high at first when $N_{r_e}$ is small but then it quickly reduces as $N_{r_e}$ increases in comparison with the case of $N_r=9$.
\section{Conclusion}
\label{sec:Conclusion}
In this work, we design an IA technique for secure multiuser MIMO system with the presence of an eavesdropper. When the CSI of main links and wiretapped links are known at Txs, the transmit precoders and the receiving subspace matrices are selected respectively in every iteration to minimize both IMLIs and wiretapped signals. Our analysis provides key insights into the proposed IA design over the proof of convergence, feasible condition and relationship between SILR and the number of antennas at Txs anf the eavesdropper. The numerical results show that the proposed IA algorithm obtained a significant SSR improvement in compared with the conventional IA method.

\balance
\bibliographystyle{IEEEtran}
\bibliography{IEEEabrv,ATC2015}

\begin{thebibliography}{10}
\providecommand{\url}[1]{#1}
\csname url@samestyle\endcsname
\providecommand{\newblock}{\relax}
\providecommand{\bibinfo}[2]{#2}
\providecommand{\BIBentrySTDinterwordspacing}{\spaceskip=0pt\relax}
\providecommand{\BIBentryALTinterwordstretchfactor}{4}
\providecommand{\BIBentryALTinterwordspacing}{\spaceskip=\fontdimen2\font plus
\BIBentryALTinterwordstretchfactor\fontdimen3\font minus
  \fontdimen4\font\relax}
\providecommand{\BIBforeignlanguage}[2]{{%
\expandafter\ifx\csname l@#1\endcsname\relax
\typeout{** WARNING: IEEEtran.bst: No hyphenation pattern has been}%
\typeout{** loaded for the language `#1'. Using the pattern for}%
\typeout{** the default language instead.}%
\else
\language=\csname l@#1\endcsname
\fi
#2}}
\providecommand{\BIBdecl}{\relax}
\BIBdecl

\bibitem{Hanif14}
M.~Hanif, L.-N. Tran, M.~Juntti, and S.~Glisic, ``On linear precoding
  strategies for secrecy rate maximization in multiuser multiantenna wireless
  networks,'' \emph{IEEE Trans. Signal Process.}, vol.~62, no.~14, pp.
  3536--3551, July 2014.

\bibitem{Yang14}
N.~Yang, G.~Geraci, J.~Yuan, and R.~Malaney, ``Confidential broadcasting via
  linear precoding in non-homogeneous {MIMO} multiuser networks,'' \emph{IEEE
  Trans. Commun,}, vol.~62, no.~7, pp. 2515--2530, July 2014.

\bibitem{Geraci12}
G.~Geraci, M.~Egan, J.~Yuan, A.~Razi, and I.~Collings, ``Secrecy sum-rates for
  multi-user {MIMO} regularized channel inversion precoding,'' \emph{IEEE
  Trans. Commun.}, vol.~60, no.~11, pp. 3472--3482, November 2012.

\bibitem{Yang13}
J.~Yang, I.-M. Kim, and D.~I. Kim, ``Optimal cooperative jamming for multiuser
  broadcast channel with multiple eavesdroppers,'' \emph{IEEE Trans. Wireless
  Commun.}, vol.~12, no.~6, pp. 2840--2852, June 2013.

\bibitem{Mukherjee09}
A.~Mukherjee and A.~Swindlehurst, ``Utility of beamforming strategies for
  secrecy in multiuser {MIMO} wiretap channels,'' in \emph{Proc. IEEE. 47th
  Annual Allerton Conf. Commun. Control and Computing}, NJ, USA, Sept 2009, pp.
  1134--1141.

\bibitem{Koyl11}
O.~Koyluoglu, H.~El~Gamal, L.~Lai, and H.~Poor, ``Interference alignment for
  secrecy,'' \emph{IEEE Trans. Inform. Theory.}, vol.~57, no.~6, pp.
  3323--3332, Jun. 2011.

\bibitem{Sasaki12}
S.~Sasaki, T.~Shimizu, H.~Iwai, and H.~Sasaoka, ``Secure communications using
  interference alignment in {MIMO} interference channels,'' in \emph{Proc. IEEE
  Int. Symp. Antenna. Propagat. (ISAP)}, Oct 2012, pp. 762--765.

\bibitem{cadambe08}
V.~R. Cadambe and S.~A. Jafar, ``Interference alignment and degrees of freedom
  of the-user interference channel,'' \emph{IEEE Trans. Inform. Theory},
  vol.~54, no.~8, pp. 3425--3441, 2008.

\bibitem{kumar10}
K.~R. Kumar and F.~Xue, ``An iterative algorithm for joint signal and
  interference alignment,'' in \emph{IEEE Int. Symp. Inform. Theory}, no.
  EPFL-CONF-172017, NJ, USA, 2010, pp. 2293--2297.

\bibitem{peters09}
S.~W. Peters and R.~W. Heath, ``Interference alignment via alternating
  minimization,'' in \emph{Proc. IEEE Int. Conf. Acoust. Speech. Signal
  Process. (ICASSP)}, Taipeo, Taiwan, Apr. 2009, pp. 2445--2448.

\bibitem{Bazz12}
S.~Bazzi, G.~Dietl, and W.~Utschick, ``Interference alignment via minimization
  projector distances of interfering subspaces,'' in \emph{Proc. IEEE Int.
  Conf. Signal Process. Advances in Wireless Commun. (SPAWC)}, Cesme, Turkey,
  Jun. 2012, pp. 274--287.

\bibitem{Lutke97}
H.~L\"{u}tkepohl, \emph{Handbook of Matrices}.\hskip 1em plus 0.5em minus
  0.4em\relax Wiley, 1997.

\bibitem{yetis09}
C.~M. Yetis, T.~Gou, S.~A. Jafar, and A.~H. Kayran, ``Feasibility conditions
  for interference alignment,'' in \emph{Proc. IEEE Conf. Global Telecommun.
  (GLOBECOM)}, Hawaii, USA, Dec. 2009, pp. 1--6.

\end{thebibliography}
\end{document}